\documentclass[pdflatex,sn-mathphys]{sn-jnl}

\usepackage{amsfonts}

\usepackage{physics}

\newcommand{\PT}{\mathcal{PT}}
\newcommand{\Psym}{\mathcal{P}}
\newcommand{\Tsym}{\mathcal{T}}
\newcommand{\llangle}{\langle\langle}
\newcommand{\rrangle}{\rangle\rangle}

\begin{document}

\title{A Brief History of Free Parafermions}


\author*{\fnm{Murray T.} 
\sur{Batchelor}}\email{murray.batchelor@anu.edu.au}

\author{\fnm{Robert A.} \sur{Henry}}

\author{\fnm{Xilin} \sur{Lu}}

\affil[]{\orgdiv{Mathematical Sciences Institute}, \orgname{Australian National University}, \orgaddress{
		\city{Canberra}, 
		\state{ACT 2601}, \country{Australia}}}


\abstract{In this article we outline the historical development and key results obtained to date for free parafermionic spin chains. The concept of free parafermions provides a natural {\it N}-state generalization of free fermions, which have long underpinned the exact solution and application of widely studied quantum spin chains and their classical counterparts. In particular, we discuss the Baxter-Fendley free parafermionic {\it Z(N)} spin chain, which is a relatively simple non-Hermitian generalization of the Ising model.
}




\maketitle

\section{Introduction}\label{sec1}

\begin{figure}[t]
	\centering
	\includegraphics[width=0.9\textwidth]{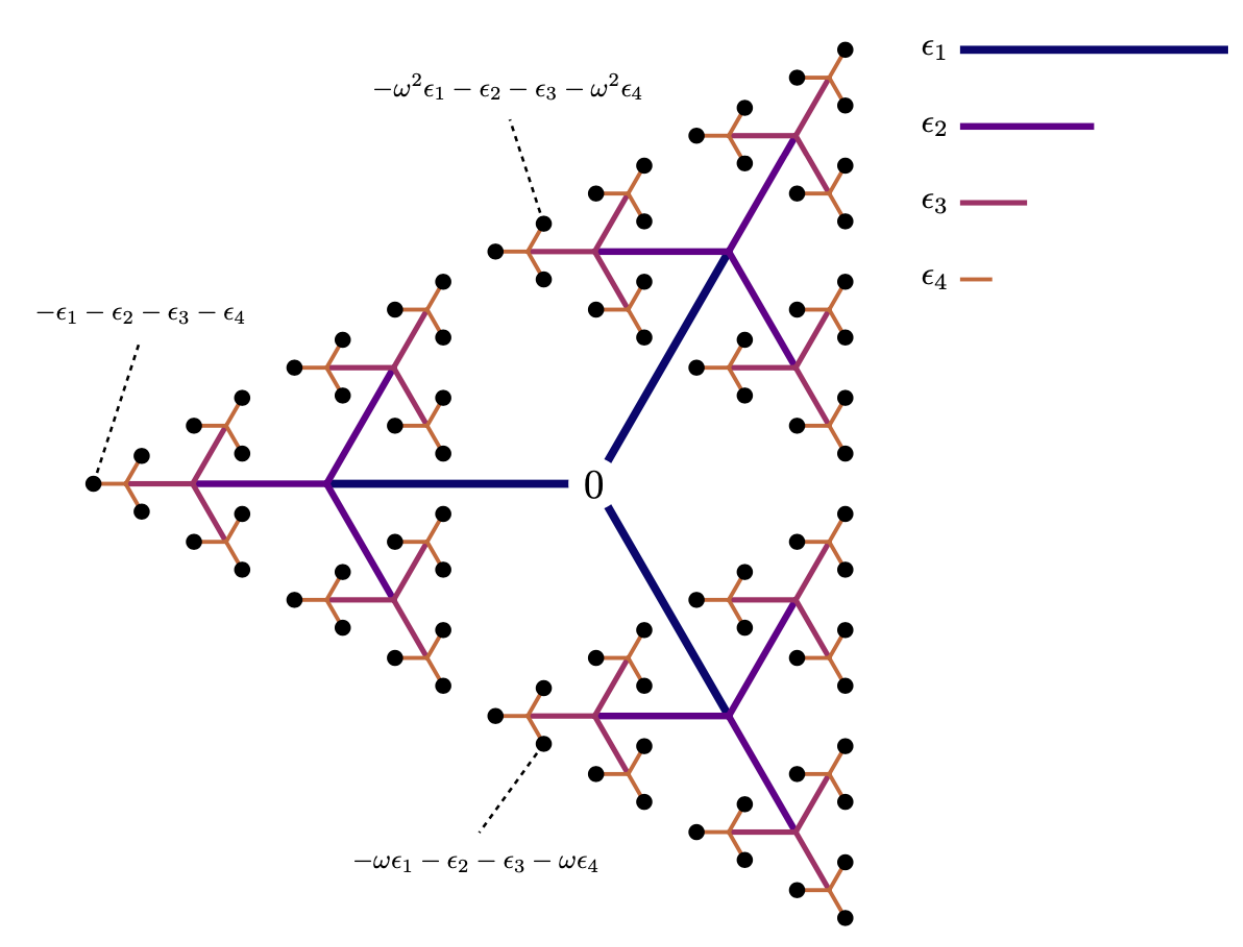}
	\caption{A free parafermion spectrum in the complex plane, for $N=3$, $L=4$. The black dots are the $N^L$ energy eigenstates. The spectrum is built up by starting at zero and adding each parafermion multiplied by a power of the root of unity $\omega = \exp(2 \pi \mathrm{i} / N)$, as per Eq.~(\ref{eq:FP_spectrum}). Here the values of $\epsilon_j$ are chosen arbitrarily and for most realistic values the paths will overlap each other, but will have the same essential branching structure. Algebraic expressions are shown for some example states, including the ground state $-\epsilon_1-\epsilon_2-\epsilon_3-\epsilon_4$.}
	\label{fig:fp_spectrum}
\end{figure}

The concept of free fermions is fundamental to the celebrated exact solution of the two-dimensional Ising model in zero magnetic field~\cite{Onsager,Kaufman1949,Lieb1964} and its one-dimensional quantum counterpart~\cite{Pfeuty}. Much later, it was revealed~\cite{Yang} that in solving the Ising model Onsager initially diagonalised the associated row transfer matrix by hand: first for strip width $L=2$,
then $L=3$, and so on. Eventually, by the $L=6$ case, he confirmed that the $2^6=64$ transfer matrix eigenvalues were
all of the form $\exp(\pm \gamma_1 \pm \cdots \pm \gamma_6)$.
This observation suggested an underlying product algebra which, in turn, led Onsager to the mathematical structure underlying his original derivation.
Insights obtained from the Ising model \textendash~specifically the exact results \textendash~played a central role in the development of the theory of phase transitions and critical phenomena.
The concept of free fermions also underpins the exact solution of many other fundamental models, with the XY chain playing an early key role~\cite{Lieb1961}.

\begin{figure}[t]
	\centering
	\includegraphics[width=0.9\textwidth]{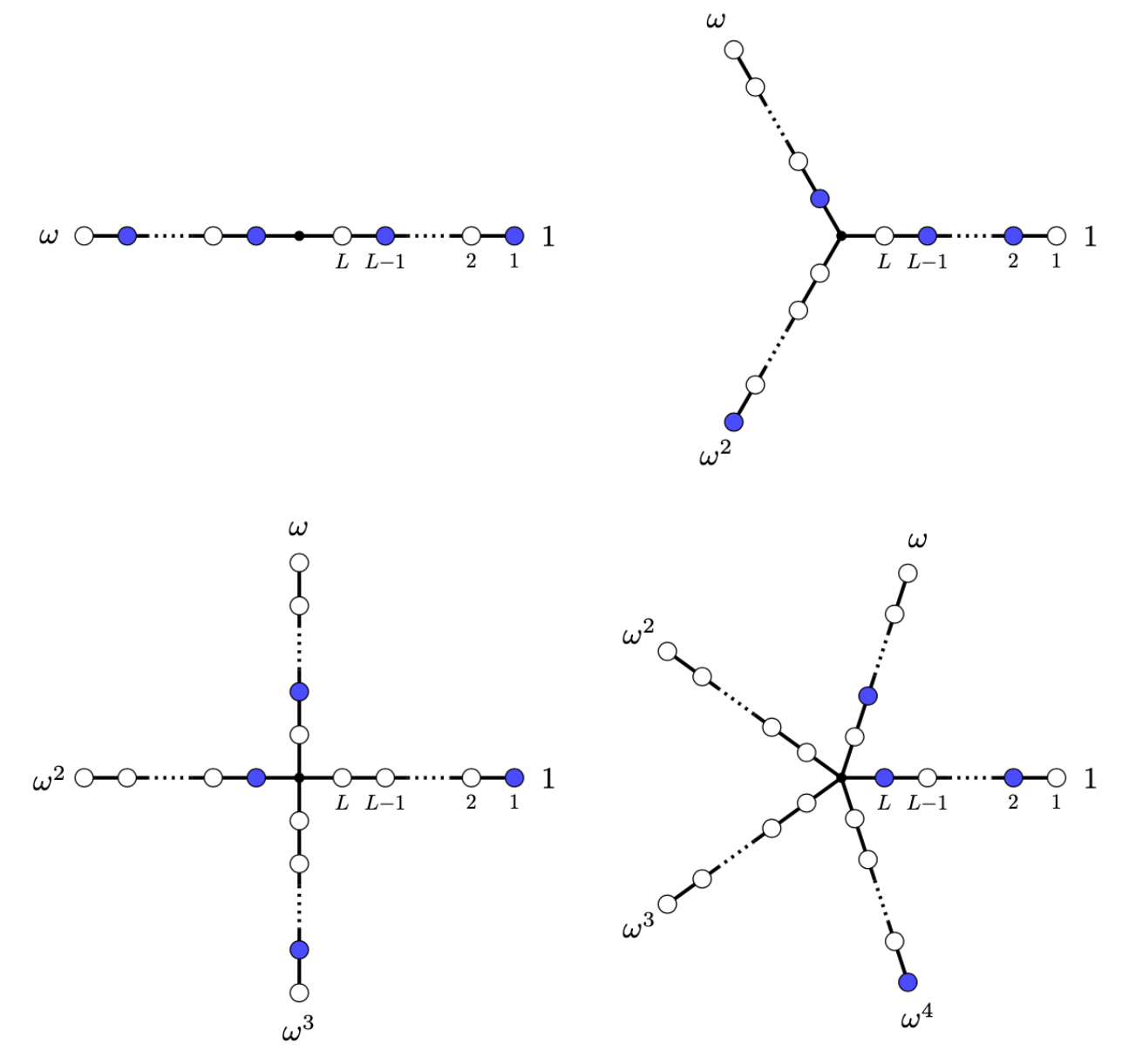}
	\caption{Free parafermion quasi-energy levels for $N=2$, $3$, $4$, $5$. Each diagram represents a single energy eigenstate of the Hamiltonian. For each $j\in\{1,\ldots,L\}$, one power of $\omega = \exp(2 \pi \mathrm{i} / N)$ is chosen. In other words, each quasi-energy is subject to a Fermi ``exclusion circle''. This gives $N^L$ possible choices, determining all the energy eigenvalues.}
	\label{fig:fp_arms}
\end{figure}

In 1989, Baxter~\cite{Baxter1989,Baxter1989chiral} derived a simple spin chain Hamiltonian from a 2D classical model known as the $\tau_2$ model, a model which was also essential to the solution of the integrable chiral Potts model~\cite{bazhanov1990chiral,Baxter2004}. As we shall see below, the Hamiltonian of this spin chain takes a similar form to that of the quantum Ising chain, but the spins have $N$ allowed states, with Baxter's Hamiltonian reducing  to the Ising model for $N=2$. Baxter found that the energy spectrum of this more general model decomposes into a sum of independent terms involving powers of $\omega$, an $N$th root of unity~\cite{Baxter1989}.

In 2012, Fendley~\cite{Fendley2012} studied the edge modes of a chiral Hermitian $Z(N)$ spin chain by rewriting the Hamiltonian in terms of parafermions. Soon after, he realised Baxter's non-Hermitian model has algebraic properties which generalise those of the fermion representation of the Ising model, and other free fermionic models. This lead to a rigorous derivation of the spectrum,  charges and the corresponding generalised Clifford algebra in 2014~\cite{Fendley2014}. This was the pivotal observation of the previously elusive free parafermions.

In this article we briefly outline related developments for free parafermions, along with further recent progress.

\section{The Free Parafermion Model}
\label{sec:FP_model}

Baxter's free parafermion model is defined by the simple Hamiltonian
\begin{equation}
	\label{eq:H_fp}
	H=-\sum_{j=1}^{L-1} Z_j^\dagger Z_{j+1} - \lambda \sum_{j=1}^L X_j,
\end{equation}
with $N$ states per site. $X$ and $Z$ are $N$-state generalisations of the Pauli matrices given by the $N\times N$ matrices
\setlength\arraycolsep{3pt}
\begin{equation}
	\label{eq:1}
	Z = \begin{pmatrix}
		1      & 0      & 0        & \ldots & 0            \\
		0      & \omega & 0        & \ldots & 0            \\
		0      & 0      & \omega^2 & \ldots & 0            \\
		\vdots & \vdots & \vdots   & \ddots & \vdots       \\
		0      & 0      & 0        & \ldots & \omega^{N-1} \\
	\end{pmatrix},\quad X = \begin{pmatrix}
		0      & 0      & 0      & \ldots & 0      & 1      \\
		1      & 0      & 0      & \ldots & 0      & 0      \\
		0      & 1      & 0      & \ldots & 0      & 0      \\
		\vdots & \vdots & \vdots & \ddots & \vdots & \vdots \\
		0      & 0      & 0      & \ldots & 1      & 0      \\
	\end{pmatrix},
\end{equation}
with $\omega = \exp(2\pi \mathrm{i}/N)$ a root of unity. The Hamiltonian is non-Hermitian and has complex eigenvalues. Baxter found that these eigenvalues have a simple form that generalises free fermions:
\begin{equation}
	\label{eq:FP_spectrum}
	E=\sum_{k=1}^L \omega^{s_k} \epsilon_k .
\end{equation}
Here $\epsilon_k$ are quasienergies which depend on $\lambda$, and the set of integers $s_k \in \{0,\ldots,N-1\}$ label each state. The $N$ choices for each $s_k$ generalise the two choices ($+1$ and $-1$, corresponding to particles and holes) in a free fermion model. This gives the model its name of \textit{free parafermions}, although this was only coined later by Fendley~\cite{Fendley2014}. Figure~\ref{fig:fp_spectrum} shows an example of a free parafermionic energy spectrum in the complex plane, subject to the generalised exclusion rule (Figure~\ref{fig:fp_arms}).

\subsection{$\PT$~Symmetry}
Although the Hamiltonian is not Hermitian, it is parity-time ($\PT$)-symmetric, with the action of the operators $\Psym$ and $\Tsym$ such that
\begin{eqnarray}
	\label{eq:PT}
	\Psym Z_j \Psym = Z_{L+1-j} \,,& & \;\Psym X_j \Psym = X_{L+1-j},\\
	\quad \Tsym Z_j \Tsym = Z_j^\dagger \,,& & \;\Tsym X_j\Tsym = X_j.
\end{eqnarray}
Bender and Boettcher showed that $\PT$-symmetric models may have real spectra~\cite{benderRealSpectraNonHermitian1998} if the symmetry is unbroken. They can be given unitary time evolution with the appropriate choice of metric, even with a broken symmetry and complex eigenvalues~\cite{mannheim2013pt}. This has lead to a great deal of activity including many experiments with $\PT$-symmetric systems, many of which are detailed in the extensive review by Ashida et al.~\cite{ashidaNonHermitianPhysics2020}. In the case of the free parafermion model, $\PT$ symmetry is always broken, and the spectrum appears in complex conjugate pairs, with the conjugation corresponding to the action of the symmetry operator.

For a finite diagonalisable system, the metric
\begin{equation}\label{eq:metric1}
 G = \sum_i \lvert L_i \rangle\langle L_i \rvert
\end{equation}
is sufficient, where $\lvert L_i\rangle$ is the $i$th left eigenstate. An inner product between two states $\lvert \Psi_1 \rangle$ and $\lvert \Psi_2 \rangle$ is then evaluated as $\langle \Psi_1 \lvert G \rvert \Psi_2 \rangle$. The expectation values of an eigenstate are easily expressed with this metric: for a right eigenstate $\lvert R\rangle$ and corresponding left eigenstate $\lvert L\rangle$ and operator $A$, the expectation value is
\begin{equation}
    \llangle A \rrangle = \langle L \rvert A \lvert R \rangle.
\end{equation}
Because the model is non-Hermitian, it is possible that the Hamiltonian is non-diagonalisable for some values of its parameters, known as {\it exceptional points}. These points have interesting physical properties and are discussed in detail in Ashida's review~\cite{ashidaNonHermitianPhysics2020}. The behaviour of the metric as the system passes through an exceptional point is the subject of current research~\cite{znojilConfluencesExceptionalPoints2022, znojilPassageExceptionalPoint2020}. At least in the case of the free parafermion model, the exceptional points are isolated, and Equation~(\ref{eq:metric1}) is usable for most calculations. Recent work on the model's exceptional points is discussed in Section~{\ref{sec13}}.

\section{From fermions to parafermions}

\subsection{The quantum Ising chain and fermions}

The widely studied Hamiltonian of the quantum Ising model on a chain of length $L$ is defined in terms of Pauli matrices by
\begin{equation}
	H_\mathrm{Ising}= -\sum_{j=1}^{L-1} \sigma^z_j \sigma^z_{j+1} - \lambda \sum_{j=1}^L \sigma^x_j, \label{IsingHam}
\end{equation}
where the subscripts indicate on which lattice site an operator is acting. The spin-spin interaction strength is scaled to unity for simplicity.
For convenience we have defined the model with open boundary conditions in contrast to period boundary conditions for which $\sigma_{L+1}^z=\sigma_1^z$.
Hamiltonian \eqref{IsingHam} possesses a $Z(2)$ symmetry,  $[H_\mathrm{Ising},(-1)^F]=0$, with
\begin{equation}
	(-1)^F = \prod_{j=1}^L \sigma^x_j.
\end{equation}

The spin operators can be rewritten in terms of fermionic operators using the Jordan-Wigner transformation. To be specific, we define the fermionic operators \begin{equation}
	\phi_{2j-1}=\prod_{k=1}^{j-1}\sigma^x_k \sigma^z_j, \quad \phi_{2j}=\prod_{k=1}^{j-1}\sigma^x_k \sigma^y_j.
\end{equation}
As fermionic operators, these operators satisfy the Clifford algebra and anticommute with each other. Moreover, they are Hermitian and square to the identity, so they are actually Majorana fermionic operators.

One can check that bilinears in neighbouring fermionic operators results in $\sigma^z_j \sigma^z_{j+1}$ and $\sigma^x_j$ for some $j$. Then it is possible to rewrite the Hamiltonian \eqref{IsingHam} as
\begin{equation}
	H=\mathrm{i}\sum_{j=1}^{2L-1} t_j \phi_j \phi_{j+1}, \label{IsingFerHam}
\end{equation}
with $t_{2j-1}=\lambda$ and $t_{2j}=1$. The key here is that Hamiltonians expressed as bilinears in fermionic operators can be diagonalised exactly by choosing an appropriate basis, of which an early example is shown in Appendix A of Ref.~\cite{Lieb1961}. The total energy is then the sum of individual fermions without any interactions in this new basis. These fermions are free and therefore such Hamiltonians are free fermionic.

\subsection{The $Z(N)$ spin chain and parafermions}
As mentioned in Section~\ref{sec:FP_model}, the free parafermion model is defined using the operators $X$ and $Z$ which generalise the Pauli matrices. A generalisation of the $Y$ operator may also be defined as
\begin{equation}
	\label{eq:4}
	Y_{i,j} =\omega^{(N-1)/2}(\delta_{N,1}+\omega^{N-i}\delta_{i+1,j})=\omega^{(N-1)/2}\begin{pmatrix}
		0      & \omega^{N-1} & 0            & \dots  & 0      \\
		0      & 0            & \omega^{N-2} & \dots  & 0      \\
		\vdots & \vdots       & \vdots       & \ddots & \vdots \\
		0      & 0            & 0            & \dots  & \omega \\
		1      & 0            & 0            & \dots  & 0
	\end{pmatrix}_{i,j},
\end{equation}

These operators satisfy the commutation relations
\begin{equation}\label{CommutationXYZ}
	XY=\omega YX,\quad YZ=\omega ZY, \quad ZX=\omega XZ,
\end{equation}
and have the property that their $N$-th power is equal to the identity operator:
\begin{equation}\label{NthPowerOfXYZ}
	A^N=\mathbb{I}, \quad A=X,Y,Z.
\end{equation}
This implies $A^{-1}=A^\dagger=A^{N-1}$ for $A=X,Y,Z$.
Also, products of two different generalised Pauli matrices in the correct order gives $\omega^{(N-1)/2}$ times the conjugate transpose of the remaining one:
\begin{equation}
	XY= \omega^{(N-1)/2}Z^\dagger,\quad YZ=\omega^{(N-1)/2} X^\dagger,\quad ZX=\omega^{(N-1)/2}Y^\dagger.
\end{equation}

Such matrices have a long history and were defined by Sylvester in 1882 as generalisations of quaternions~\cite{parshall1998james}. Like quaternions, they satisfy a (generalised) Clifford algebra defined by the above commutation relations. 
This leads to a set of parafermion generators, or simply ``parafermions", which have appeared in various forms and were studied by Yamazaki~\cite{Yamazaki1964} and Morris~\cite{Morris1967} in the context of the  generalised Clifford algebra. For other early related references to parafermions the reader is referred to Jaffe and Pedrocchi~\cite{Jaffe2015}, who investigated the relation with reflection positivity, and also Fradkin and Kadanoff~\cite{FK1980} who introduced the $Z(N)$ generalisation of the Jordan-Wigner transformation.

The free parafermion model \eqref{eq:H_fp} can be obtained from the Ising chain by directly substituting the generalised operators.
It is straightforward to check that \eqref{eq:H_fp} is invariant under the $Z(N)$ symmetry operator \begin{equation} \label{ZNOperator}
	\omega^\mathcal{P}\equiv \prod_{j=1}^L X_j,
\end{equation}
with $(\omega^\mathcal{P})^N=\mathbb{I}$ following from~\eqref{NthPowerOfXYZ}. Following Fendley's work \cite{Fendley2014}, we can rewrite this Hamiltonian~\eqref{eq:H_fp} in terms of parafermionic operators $\psi_j$ acting on a lattice of length $2L$:
\begin{equation}\label{ParaferDef}
	\psi_{2j-1}=\prod_{k=1}^{j-1}X_k Z_j, \quad \psi_{2j}=\omega^{-1}\prod_{k=1}^{j-1}X_k Y_j^\dagger.
\end{equation}
These operators satisfy $\psi_j^N = 1$, and the  $\omega$-commutation relation 
\begin{equation}
	\psi_a \psi_b=\omega \psi_b \psi_a, \quad a<b.
\end{equation}
This can be verified with~\eqref{NthPowerOfXYZ} and~\eqref{CommutationXYZ}. In fact, Fendley's analysis extrends to any nonuniform couplings $t_j$. These operators are {\it parafermions} and generalise Majorana fermions to the $Z(N)$ case. It is not hard to check that the Hamiltonian \eqref{eq:H_fp} is a sum of bilinears in parafermionic operators:\begin{equation}
	H=-\omega^{-\frac{N-1}{2}}\sum_{j=1}^{2L-1} t_j \psi_j^\dagger \psi_{j+1}, \label{ZNParaHam}
\end{equation}
with $t_{2j-1}=\lambda$ and $t_{2j}=1$. Starting with this form, Fendley~\cite{Fendley2014} performs a series of linear transformations which determine the free spectrum and eigenvectors. Although $H$ is expressed as a sum of bilinears of parafermions, this does not guarantee that it has a free spectrum, unlike the fermionic case~\cite{stoudenmire2015assembling}. A general condition for free spectra is given by the exchange algebra discussed in Section~\ref{sec:exchange_algebra}.

\subsection{Spectrum and other physical quantities}

Baxter's method \cite{Baxter1989,Baxter1989chiral} to calculate the spectrum of \eqref{eq:H_fp} can be summarised as follows. Analogously to the Ising case, we interlace the coupling constants $1$ and $\lambda$ to define the $2L\times 2L$ symmetric matrix $B$:
\begin{equation}
	\label{eq:quasienergy_matrix}
	B= \begin{pmatrix}
		0             & \lambda^{N/2} &                                                          \\
		\lambda^{N/2} & 0             & 1                                                        \\
		              & 1             & 0      & \ \lambda^{N/2}                                 \\
		              &               & \ddots & \ddots          & \ddots                        \\
		              &               &        & 1               & 0             & \lambda^{N/2} \\
		              &               &        &                 & \lambda^{N/2} & 0
	\end{pmatrix},
\end{equation}
where the powers $\lambda^2$ found in the Ising case are replaced by $\lambda^{N/2}$. The off-diagonal entries are the coefficients $t_j^{N/2}$ in~\eqref{ZNParaHam}, so the matrix $B$ represents the Hamiltonian in the parafermion basis.
The next step is calculating the quasienergies $\epsilon_k$ as the solutions of the equation \begin{equation}
	\det{\epsilon^{N/2} \mathbb{I}-B}=0.
\end{equation}

This equation is of degree $L$ in $\epsilon^{N}$. From the $Z(N)$ symmetry, we know the eigenvalues are $N$-fold degenerate, thus only $L$ independent $\epsilon_k$'s are left. This implies the form of the free spectrum, as given in \eqref{eq:FP_spectrum}.
Baxter's argument is based on analogies to the $N=2$ Ising case and does not give any information beyond the eigenvalues. Fendley's derivation of the free spectrum~\cite{Fendley2014} also determines the eigenvectors in terms of the parafermion operators, as well as other aspects such as conserved charges and higher Hamiltonians.

\section{The Exchange Algebra}\label{sec:exchange_algebra}

Fendley's solution~\cite{Fendley2014} makes use of the fact that the terms in the Hamiltonian obey a simple exchange algebra. He uses a similar algebra in related work~\cite{fendleyFreeFermionsDisguise2019} on free fermions. These are both forms of a more general algebra identified by Alcaraz and Pimenta~\cite{Alcaraz2020}, which is satisfied by a large class of free parafermionic models. This algebra is defined for a general Hamiltonian of the form
\begin{equation}
	\label{eq:2}
	H = \sum_{i=1}^M h_i \,.
\end{equation}
The $M$ generators $h_i$ satisfy the algebra
\begin{align}
	\label{ExchangeAlgebra}
	  h_i h_{i+m} &= \omega \, h_{i+m}h_i \quad \mathrm{for}\quad 1\leq m\leq p, \nonumber \\
	  [h_i, h_j] &= 0  \qquad\qquad ~ \, {\mathrm{for}}\quad \lvert i-j \rvert > p,
\end{align}
and the closure relation $h_i^N = \lambda_i^N$, 
where $\omega$ and $\lambda_i$ are complex numbers. The parameter $p$ is a positive integer determining the range of the ``interaction'' between terms within which they $\omega$-commute, i.e.,~ $p > 1$ is a multispin model. Alcaraz and Pimenta show that any Hamiltonian obeying this algebra is integrable and has a free parafermion spectrum of the form \eqref{eq:FP_spectrum}, for some values of the parafermion quasienergies $\epsilon_j$. The Baxter Hamiltonian \eqref{eq:H_fp} satisfies the algebra with $\omega = \exp(2 \pi \mathrm{i}/N)$ and $M=2L-1$ the total number of terms in the Hamiltonian. 

This algebra has been used to explore generalised free parafermion models, including a class of multispin $XY$-type models  with spectra composed of combinations of free parafermions~\cite{Alcaraz2020multispin, Alcaraz2021}. It has also been used to develop an efficient numerical method for calculating the mass gaps associated with these free particle modes~\cite{Alcaraz2021gap}.
Most recently, it has been shown~\cite{Alcaraz2023} how to build standard quantum Ising chains with inhomogeneous couplings which have the same spectra as the new family of free fermionic quantum spin chains with multispin interactions.

 Minami~\cite{minamiOnsagerAlgebraAlgebraic2021} has provided a general list of models composed of the $X$, $Y$ and $Z$ operators which satisfy generalised Onsager algebras.

\subsection{Polynomial Expressions}
Alcaraz and Pimenta~\cite{Alcaraz2020}  show that the quasienergies $\epsilon_j$ for a general Hamiltonian satisfying the algebra~\eqref{ExchangeAlgebra} are given by the roots $z_i$ of a polynomial $P_M^{(p)}(z)$, with $\epsilon_i = z_i^{-1/N}$. This provides a practical way to determine the quasienergies for a general model satisfying the algebra, although it is less efficient than diagonalising~\eqref{eq:quasienergy_matrix}. There are $\bar{M} = \lfloor\frac{M+p}{p+1}\rfloor$ quasienergies, and the polynomial is determined by the recursion relation
\begin{align}
	\label{eq:pf_polys}
	P_M^{(p)}(z) = \sum_{l=0}^{\bar{M}} C_M(l) z^l,
\end{align}
which satisfies a recurrence relation for $M\geq$ 1
\begin{align}
	P_M^{(p)}(z) = P^{(p)}_{M-1}(z) - z\lambda_M^N P_{M-(p+1)}^{(p)}(z),
\end{align}
and an initial condition $P^{(p)}_M(z) = 1$ for $M \leq 0$.


\section{Other Developments}\label{sec13}

Before turning to other developments, we remark that
Hamiltonians obeying relations similar to the exchange algebra with $\omega=-1$ have been shown~\cite{Elman2021} to have explicit free fermionic spectra.
More specifically, frustration graphs are drawn for Hamiltonians representing their local Hamiltonian commutation relations. If the resulting graph of a Hamiltonian is free of ``even-hole'' or ``claw'' structures, then a free fermion spectrum can be constructed. It would be worthwhile attempting this approach with $Z(N)$ clock models.

Much is known, including zero mode criticality and low energy CFT~\cite{Li2015criticality}, about the free parafermion model when Hermitian conjugate (h.c.) terms are added to the Hamiltonian.
For example, including the h.c. terms for $N=3$ results in the well known 3-state Potts model. So far various critical properties have been calculated for the free parafermion Hamiltonian \eqref{eq:H_fp}. These include the bulk ground state energy and critical exponents~\cite{Alcaraz2017} and certain correlations~\cite{Liu2019}. Insights from the free parafermionic structure of the $Z(N)$ spin chain were also used to study the $\tau_2$ model \cite{Baxter2014,Perk2014,Perk2016}.

It should be stressed that the free parafermion model has only been solved for open boundary conditions. An unexpected numerical observation is that the ground state energy depends on the boundary conditions in the bulk ($L\to\infty$) limit~\cite{Alcaraz2018}. This is a characteristic property of non-Hermitian systems, likely being  an example of a non-Hermitian skin effect~\cite{bergholtzExceptionalTopologyNonHermitian2021}, which can only occur in a non-Hermitian system and is caused by macroscopic occupation of boundary states.

By extending $\lambda$ to complex values, it has most recently been shown that a series of exceptional points appears where two quasienergies become degenerate~\cite{henry2023exceptional}. This leads to a macroscopic number of degeneracies in the eigenvectors of the full Hamiltonian. Exceptional points are particular to non-Hermitian systems and have many interesting properties~\cite{ashidaNonHermitianPhysics2020}. These exceptional points are also seen to appear in the $N=2$ Ising case for complex values of $\lambda$.

\section*{}

\bmhead{Acknowledgements}
See funding support.

\bmhead{Availability of data and materials}
N/A

\bmhead{Competing interests} 
The authors declare they have no competing interests.

\bmhead{Authors' contributions}
The authors have contributed equally to this manuscript.

\bmhead{Funding}
This work has been supported by the Australian Research Council's Discovery Program through grant number DP210102243.



\end{document}